# Intrusion Detection and identification System Design and Performance Evaluation for Industrial SCADA Networks

Ahsan Al Zaki Khan and Gursel Serpen

**Abstract**

In this paper, we present a study that proposes a three-stage classifier model which employs a machine learning algorithm to develop an intrusion detection and identification system for tens of different types of attacks against industrial SCADA networks. The machine learning classifier is trained and tested on the data generated using the laboratory prototype of a gas pipeline SCADA network. The dataset consists of three attack groups and seven different attack classes or categories. The same dataset further provides signatures of 35 different types of sub-attacks which are related to those seven attack classes. The study entailed the design of three-stage machine learning classifier as a misuse intrusion detection system to detect and identify specifically each of the 35 attack subclasses. The first stage of the classifier decides if a record is associated with normal operation or an attack signature. If the record is found to belong to an attack signature, then in the second stage, it is classified into one of seven attack classes. Based on the identified attack class as determined by the output from the second stage classifier, the attack record is provided for a third stage sub-attack classification, where seven different classifiers are employed. The output from the third stage classifier identifies the sub-attack type to which the record belongs. Simulation results indicate that designs exploring specialization to domains or executing the classification in multiple stages versus single-stage designs are promising for problems where there are tens of classes. Comparison with studies in the literature also indicated that the multi-stage classifier performed markedly better.

# Introduction

Supervisory Control and Data Acquisition (SCADA) systems monitor and control highly critical industrial infrastructure. Such systems gather and analyze data, and control processes and systems all in real time for the most part. SCADA systems are used to monitor and control a plant or equipment such as gas pipeline, water storage tank and associated distribution network, telecommunications, waste control, oil refining and transportation among many others. A SCADA system may collect information such as where a leak on a gas pipeline has occurred; alert the central control room that leak has occurred; and carry out necessary analysis and control (such as determining if the leak is critical or not). A SCADA system can be very simple i.e. just monitoring environment of a small manufacturing facility or it can be very complex such as monitoring activity of an oil refinery or a nuclear power plant.

Computers were first used for industrial control purposes as early as late 1950s [2]. Telemetry was established for monitoring in the 1960s, which allowed for automated communications to transmit measurements. In the early 1970s, the term "SCADA" was coined and the rise of microprocessors and programmable logic controllers (PLCs) during that decade increased enterprises' ability to monitor and control automated processes more than ever before. SCADA systems have undergone significant changes in subsequent decades. During late 1990s to early 2000s, a technological revolution occurred as computing and information technologies (IT) accelerated in growth. The introduction of modern IT standards and practices such as Structured Query Language (SQL) and web-based applications for SCADA networks has improved the efficiency and productivity overall. Many SCADA systems are either online or able to connect to other similar systems or both, and with this newfound connectivity, there are also many security concerns for these once remote, isolated and standalone systems [7]. If a vulnerability exists in one of these systems, it will now allow attackers to remotely exploit and potentially be able to take control of these SCADA systems; the stakes then could not be higher as takeover by a bad actor could lead to unimaginable and catastrophic consequences. Table 1 provides some common SCADA attack scenarios. Hong and Lee discuss inherent security issues in SCADA systems for smart grid communications [6]. Similar to this work, Dzung et al. outlines many issues found in communication networks for industrial applications [4]. Mirian et al. [9] found that 60,000 vulnerable SCADA devices connected to the Internet using Internet wide scanner called Zmap [3]. A detailed survey of risk assessment studies reported in the literature for industrial SCADA systems is presented in [23].

Table 1. Common SCADA System Threats [7]

| Common SCADA Threats | | | | |
|---|---|---|---|---|
| *Sabotage* | *Scavenging* | *Spying* | *Spoofing* | *Substitution* |
| *Worm* | *Access Violation* | *Trojan Horse* | *Tunneling* | *Theft* |
| *Information Leakage* | *Data Modification* | *Physical Intrusion* | *Resource Exhaustion* | *Traffic Analysis* |
| *Eavesdropping* | *Repudiation* | *Intercept* | *Terrorism* | *Virus* |

Given that online SCADA systems are prone to cyber-attacks due their large-scale deployments and distributed network operating modes, many researchers have been trying to build Intrusion Detection Systems (IDS). Intrusion detection systems (IDS) are used to collect data and analyze system activity to monitor a system's status and state [8]. Many IDSs use machine learning algorithms for pattern recognition to detect and identify any threat activity. There are mainly two types of IDSs. One type uses a signature-based approach to compare activity to a database of known threats, and as such are considered to perform misuse detection. The other type can identify an operation mode of the system as outside the boundaries of normal mode, which are then characterized as performing anomaly detection. These functionalities can be combined for a robust detection system and will likely form a baseline design for minimally adequate layer of protection against attacks.

The highly critical operational nature of SCADA systems mandates using Intrusion Detection Systems (IDSs) for defense against attacks exploiting vulnerabilities in those systems. A recent study [1] used real world data from an industrial system (water plant) to experiment with two different approaches. The research concludes with a finding that behavioral approach for intrusion detection can help yield high detection rates for SCADA networks. Cheng et. al proposed a deep learning-based framework to detect attacks against SCADA networks in industrial systems [5]. Their framework shows that Artificial Intelligence (AI) can be helpful to detect even stealthy attacks on SCADA systems given that such attacks are normally very hard to detect. Several other industrial control system specific anomaly and intrusion detection system models have been proposed in [16] [17] [21]. Bigham et al. [18] proposed statistical Bayesian networks for SCADA systems. In [19], Nader et al. proposed one-class classification for SCADA IDS using support vector machines (SVM). Perez et. al [14] used Random Forest to build an IDS and classify attacks in gas pipeline-based SCADA systems. Simon et al. proposed anomaly-based intrusion detection with industrial data with both SVM and Random Forest [20].

Current industrial SCADA networks are facing constantly evolving threats from hackers with potentially catastrophic consequences for mission-critical tasks. The defensive tools must be also in a state of evolution to address these changing threats. New vulnerabilities are being exploited by the adversaries of such systems, which requires a constant engagement in terms of engineering such systems for defense. Therefore, there is an ongoing and urgent need to continue with development of Intrusion Detection Systems to counter the existing or future threats being posed to such systems.

SCADA networks for industrial infrastructure employ networking protocols to facilitate communication for command and control. There is much information embedded in these networking packets which can be leveraged for intrusion detection purposes. This requires collection of data to be used for development of data-driven decision-making tools such as machine learning classifiers. The significant and substantial differences in the design and architecture of SCADA networks for different industrial settings poses a hindrance to development efforts as an intrusion detection system developed for a water distribution system cannot be readily adopted for an oil refinery, gas pipeline [10] or industrial manufacturing plant.

There are two main and interrelated objectives of the research study presented in this paper. First such objective is to explore the information content of networking packets for communication and command in SCADA networks to determine the feasibility of identification of detected attacks at a multitude of levels as a) Detection of attacks occurring versus normal operation; b) Detection and identification of a specific attack type which belongs to a group occurring where a group consists of several attacks sharing common attributes; and c) Detection and identification of a specific attack occurring in a context where there are tens of such attacks can occur. The second objective is to explore the performance of a multi-stage classifier architecture design.

## Dataset Description, Preprocessing and Training-Testing Partitioning

This section presents the Gas Pipeline dataset [12], its features and attack class labels. It also presents the preprocessing steps and methods applied to the datasets to fill in the missing values.

### Dataset Features

The original dataset has 17 features and 3 different class label groups namely binary, categorized and specified. There are a total 274,628 instances in the dataset. There are 11 Command Payload features which are related to the command injection attacks, 5 Network features, and 1 Response Payload feature related to the response injection attacks. The description of features and the associated collection method can be found in Ian et al. [12]. The feature set is listed in Table 2.

### Attack Types

The gas pipeline dataset used in this study has 7 types or categories of attacks as presented in Table 3. The description for all attack types are given in Gao et al [13]. Naïve Malicious Response Injection (NMRI) and Complex Malicious Response Injection (CMRI) are the response injection attacks. These attacks can hide by mimicking certain behaviors which occur within normal operating bounds. This makes them very difficult to detect, and hence giving the appearance of the system operating normally. NMRI has out of bounds behavior that would not be present in normal operation. It typically occurs when the attacker lacks information about the physical system process. CMRI attacks provide a level of sophistication over NMRI attacks. These attacks can change states which can be seen as command injection attacks. Since these attacks inject states, they become more difficult to detect.

Table 2. Original features in gas pipeline dataset [12]

| Features | Type | Values |
| --- | --- | --- |
| *Address* | Network | Numeric |
| *Length* | Network | Numeric |
| *Gain* | Command Payload | Numeric |
| *Deadband* | Command Payload | Numeric |
| *Rate* | Command Payload | Numeric |
| *Control Scheme* | Command Payload | 0 or 1 |
| *Solenoid* | Command Payload | 0 or 1 |
| *CRC Rate* | Network | Numeric |
| *Timestamp* | Network | Numeric/Integer/UNIX Format |
| *Function* | Command Payload | Numeric |
| *Set Point* | Command Payload | Numeric |
| *Reset Rate* | Command Payload | Numeric |
| *Cycle Time* | Command Payload | Numeric |
| *System Mode* | Command Payload | 0 or 1 or 2 |
| *Pump Mode* | Command Payload | 0 or 1 |
| *Pressure Measurement* | Response Payload | Numeric |
| *Command Response* | Network | 0 or 1 |
| *Binary Attacks* | Label | 0 or 1 |
| *Categorized Attacks* | Label | 0, 1, 2…,7 |
| *Specified Attacks* | Label | 0, 1, 2…, 35 |

Table 3. Attack classes in gas pipeline dataset

| Attack Type/Category/Class Name | Acronym | Instance Count |
| --- | --- | --- |
| Normal | n/a | 214580 |
| Naïve Malicious Response Injection | NMRI | 7753 |
| Complex Malicious Response Injection | CMRI | 13035 |
| Malicious State Command Injection | MSCI | 7900 |
| Malicious Parameter Command Injection | MPCI | 20412 |
| Malicious Function Code Injection | MFCI | 4898 |
| Denial of Service | DoS | 2176 |
| Reconnaissance | Recon | 3874 |

Table 4. MPCI Attack Subtypes

| Attack Name | Class Number | Class Type | Description |
|---|---|---|---|
| Setpoint Attacks | 1-2 | MPCI | Changes the pressure set point outside and inside of the range of normal operation. |
| PID Gain Attacks | 3-4 | MPCI | Changes the gain outside and inside of the range of normal operation. |
| PID Reset Rate Attacks | 5-6 | MPCI | Changes the reset rate outside and inside of the range of normal operation. |
| PID Rate Attacks | 7-8 | MPCI | Changes the rate outside and inside of the range of normal operation. |
| PID Deadband Attacks | 9-10 | MPCI | Changes the dead band outside and inside of the range of normal operation. |
| PID Cycle Time Attacks | 11-12 | MPCI | Changes the cycle time outside and inside of the range of normal operation. |

Malicious State Command Injection (MSCI), Malicious Parameter Command Injection (MPCI), and Malicious Function Code Injection (MFCI) labels belong to the command injection attacks. Tables 4 and 5 show their specific attack types and their adverse impact on the system. Much damage may originate from command injections attacks: interruption in device communications, modification of device configuration, and modification of the PID values are some of them. MSCI attacks modify the state of the current physical process of the system and can potentially place the system into a critical state. Table 4 specifies the MPCI attack types. It mainly modifies the parameters of PID configurations and set point.

As listed in Table 5, MFCI attacks inject commands which exploit the network protocol for restarting, cleaning registers etc.

Table 5. MSCI, MFCI, DoS, Recon Attack Subtypes

| Attack Name | Class Number | Class Type | Description |
| --- | --- | --- | --- |
| Pump Attack | 13 | MSCI | Randomly changes the state of the pump. |
| Solenoid Attack | 14 | MSCI | Randomly changes the state of the solenoid. |
| System Mode Attack | 15 | MSCI | Randomly changes the system mode. |
| Critical Condition Attacks | 16-17 | MSCI | Places the system in a Critical Condition. This condition is not included in normal activity. |
| Bad CRC Attack | 18 | DoS | Sends Modbus packets with incorrect CRC values. This can cause denial of service. |
| Clean Register Attack | 19 | MFCI | Cleans registers in the slave device. |
| Device Scan Attack | 20 | Recon | Scans for all possible devices controlled by the master. |
| Force Listen Attack | 21 | MFCI | Forces the slave to only listen. |
| Restart Attack | 22 | MFCI | Restarts communication on the device. |
| Read ID Attack | 23 | Recon | Reads ID of slave device. The data about the device is not recorded but is performed as if it were being recorded. |
| Function Code Scan Attack | 24 | Recon | Scans for possible functions that are being used on the system. The data about the device is not recorded but is performed as if it were being recorded. |

Denial of service (DoS) attacks are very common in almost every networked and online system. In a SCADA system, a DoS attack attempts to disrupt communication between the control or monitoring system and the process. Another category of attacks are reconnaissance attacks. These attacks aim to collect information about the system through some passive activity. They may also query the device for information such as function codes, model numbers etc. Specific attack types belonging to NMRI or CMRI are listed in Table 6.

Table 6. NMRI & CMRI Attack Subtypes

| Attack Name | Class Number | Class Type | Description |
|---|---|---|---|
| Rise/Fall Attacks | 25-26 | CMRI | Sends back pressure readings which create trends. |
| Slope Attacks | 27-28 | CMRI | Randomly increases/decreases pressure reading by a random slope |
| Random Value Attacks | 29-31 | NMRI | Random pressure measurements are sent to the master. |
| Negative Pressure Attacks | 32 | NMRI | Sends back a negative pressure reading from the slave. |
| Fast Attacks | 33-34 | CMRI | Sends back a high set point then a low setpoint which changes "fast" |
| Slow Attack | 35 | CMRI | Sends back a high setpoint then a low setpoint which changes "slowly" |

**Preprocessing**

In the preprocessing stage, missing values in the dataset were filled in. There were missing values in the dataset for 11 Payload (10 Command Payload and 1 Response Payload) features. The missing values could have been imputed in multiple ways. For instance, Perez et. al. [14] have imputed the missing values in 4 different ways such as mean value, keeping previous value, zero imputation, and K-means imputation for the same dataset. In the gas pipeline dataset, missing values were occurring as MAR (Missing At Random) or NMAR (Not Missing At Random). The payload features were not occurring at random as the solenoid or pump mode values were fixed among 0,1 or on/off/automatic. But the pressure measurement value was occurring randomly while an associated attack was in progress. Accordingly, the missing values were imputed with Multivariate Imputation by Chained Equation (MICE) method [11] since the MICE algorithm can handle both MAR and NMAR types. This type of imputation works by filling the missing data multiple times. Multiple Imputations (MIs) are much better than a single imputation as it measures the uncertainty of the missing values more precisely [11]. The chained equations approach is also very flexible and can handle different variables or different data types.

**Dataset Partitioning for Training and Testing**

To create the train-test data subsets, we partitioned the original dataset into three splits and then formed training-testing datasets at a ratio of 66.6% and 33.3% and repeated three times to form three data folds. The original dataset was split into three equal parts while preserving class representations equally in each split. Using these 3 splits, we created 3 data folds. Each fold contains two splits for training and the third split is used for testing: this procedure generates a unique split for testing for a given fold.

## Classifier Design

The design of the 3-stage machine learning classifier is illustrated in Figure 1. For the classification algorithm in all three stages, we have chosen Random Forest [15] given its superior performance as presented in a previous publication by the authors [22]. The first stage classifier performs binary classification outputting if the class is normal or attack. If the classification output is normal, then no further action is taken. However, if the classification output is attack, then second stage classifier is activated. The pattern that was classified as attack by the first stage binary classifier is input to the second stage classifier. The second stage classifier performs attack class identification on the input pattern that was classified as an attack (vs. normal) by the first stage binary classifier.

There are 7 class labels or outputs from the second stage classifier corresponding to 7 attack classes. Once the class label is identified by the second stage classifier the corresponding third stage classifier is activated to perform subclass identification on the attack pattern. There are 7 subclass classifiers performing this task in the third stage: six of the seven are implemented noting that there is only one subclass for the DoS attack.

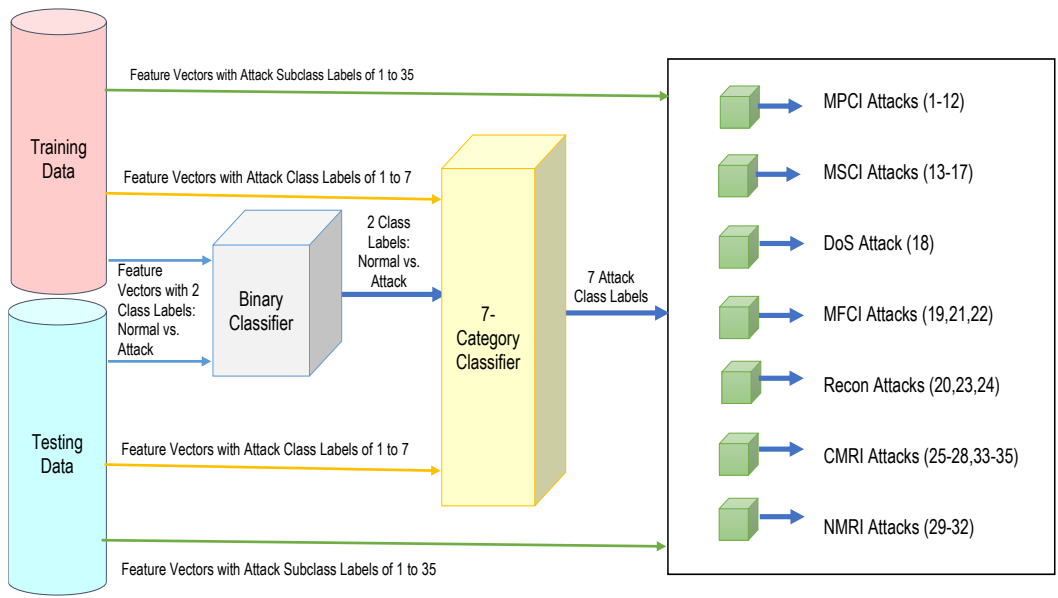

FIGURE 1. 3-STAGE CLASSIFER DESIGN

## Simulation Study

Simulation study was performed to determine the performance of the 3-stage classifier on each of the 3 folds. Performance assessment and evaluation was done separately for each test data split and further inferences were done based on these three performance assessments and evaluations.

Training the 3-stage classifier was accomplished by implementing the following procedure:

> Step 1: Train the single binary classifier in Stage 1 with two class labels of Normal versus Attack.
> Step 2: Remove all patterns with class label "Normal" and replace the single "Attack" class label with 7 attack class labels in the training dataset
> Step 3: Train the single 7-category classifier in Stage 2 with training dataset modified by removing normal class patterns and replacing the single "Attack" label with 7 attack class labels.
> Step 4: Split the single training dataset into 7 subsets (one for each attack class); replace, in each training subset, each of the 7 class labels with the corresponding subset of subclass labels. A training data subset will have those records or instances belonging to subclasses of a given attack class.
> Step5: Train each of the 7 classifiers for subclass classification using corresponding training data subsets.

Performance assessment and evaluation was done by implementing the following procedure:

> Step 1: Classify the pattern in the testing data subset with binary class labels using Stage 1 binary classifier.
> If classification output is "Normal" then no further processing is needed, then exit.
> Else (classification must be "Attack") continue with Stage 2 processing.
> Step 2: Classify the attack pattern into one of seven classes in the testing data subset using the 7-category classifier in stage 2. Based on the output from the single 7-category classifier, chose the Stage 3 classifier among 7 to activate.

Step 3: Classify the attack pattern using the subclass classifier (among those 7 at stage 3) as identified in Stage 2 for the subclass membership of attack patterns in the testing data subset.

**Simulation Results and Discussion**

Weka classifier Random Forest was used for all classifier models in all 3 stages [22]. The training set of the first data fold contains dataset splits 1 and 2 while the dataset split 3 is employed as the test set. Table 7 presents the number of instances available in both training and testing data subsets for stage 1. We report the representative results for one of the three data folds since results for all three folds are very similar to each other with negligible differences.

Table 7. Instance Counts for Stage 1

|          | **Classes** | |        |
|----------|-------------|--------|--------|
| **Dataset** | Normal  | Attack | Total  |
| Training | 142901      | 40138  | 183039 |
| Testing  | 71679       | 19910  | 91589  |

For stage 1, the batch size is 1000 for the Random Forest model which classified 89,906 instances correctly from the test set yielding 98.16% accuracy. The confusion matrix in Table 8 shows 18,597 attack instances were classified correctly among the 19,910 and thus resulting in 1,313 incorrectly classified attack instances. From Table 9, we see that false negative rate (FNR) for the attack class is 6.6% which corresponds to the 1,313 incorrectly classified attack instances.

Table 8. Stage 1 Confusion Matrix

| Classified as → | Normal | Attack |
|-----------------|--------|--------|
| Normal          | 71309  | 370    |
| Attack          | 1313   | 18597  |

Table 9. Stage 1 Performance

| Class | Accuracy | TPR | FPR | TNR | FNR | Precision | Recall |
|---|---|---|---|---|---|---|---|
| **Normal** |  | 0.995 | 0.066 | 0.934 | 0.005 | 0.982 | 0.995 |
| **Attack** |  | 0.934 | 0.005 | 0.955 | 0.066 | 0.980 | 0.934 |
| **Weighted** | 98.16% | 0.982 | 0.053 | 0.947 | 0.018 | 0.982 | 0.982 |

Next, the training set is modified: all 142,901 normal class patterns are removed from the training set of stage 1. This leaves only the attack class instances in the training set for stage 2. Table 10 presents the number of instances available in training and testing data subsets for stage 2.

Table 10. Instance Counts for Stage 2

| Class Label | 1 | 2 | 3 | 4 | 5 | 6 | 7 | Total |
|---|---|---|---|---|---|---|---|---|
| **Training Count** | 5222 | 8743 | 5361 | 13550 | 3232 | 1449 | 2581 | 40138 |
| **Testing Count** | 2531 | 4292 | 2539 | 6862 | 1666 | 727 | 1293 | 19910 |

The batch size used for stage 2 classification is 100 as the number of instances decreased in comparison to stage 1. Random Forest at Stage 2 classifies with 93.79% accuracy where 18,674 test instances are correctly classified and 1,236 are incorrectly classified. The weighted FNR is 6.2% for stage 2. From the confusion matrix in Table 11 we see that the model mainly struggles distinguishing between attack classes 1 and 2. There are 1161 incorrectly classified instances between attack classes 1 and 2 which amounts to approximately 94% of incorrectly classified instances. NMRI and CMRI, being both response injection attacks, deal with mainly the pressure measurement features which have overlapping values for both attack classes 1 and 2. For example, both attacks have same pressure values in the range of 2 to 10 kPa. That is the most likely reason why the classifier struggles to distinguish between them. This finding also indicates the need to determine and formulate new features which would be discriminatory between attack classes 1 and 2.

Table 11. Stage 2 Confusion Matrix

| Classified as → | 1 | 2 | 3 | 4 | 5 | 6 | 7 |
|---|---|---|---|---|---|---|---|
| 1 | 1881 | 650 | 0 | 0 | 0 | 0 | 0 |
| 2 | 511 | 3781 | 0 | 0 | 0 | 0 | 0 |
| 3 | 0 | 0 | 2522 | 15 | 0 | 2 | 0 |
| 4 | 0 | 0 | 22 | 6835 | 0 | 5 | 0 |
| 5 | 0 | 0 | 0 | 0 | 1666 | 0 | 0 |
| 6 | 0 | 0 | 6 | 15 | 0 | 706 | 0 |
| 7 | 0 | 0 | 0 | 0 | 10 | 0 | 1283 |

Table 12. Stage 2 Performance

| Class | Accuracy | TP Rate | FP Rate | TN Rate | FN Rate | Precision | Recall |
|---|---|---|---|---|---|---|---|
| 1 |  | 0.743 | 0.029 | 0.971 | 0.257 | 0.743 | 0.764 |
| 2 |  | 0.881 | 0.042 | 0.958 | 0.190 | 0.853 | 0.881 |
| 3 |  | 0.993 | 0.002 | 0.998 | 0.007 | 0.989 | 0.993 |
| 4 |  | 0.996 | 0.002 | 0.998 | 0.004 | 0.996 | 0.996 |
| 5 | 93.79% | 1.000 | 0.001 | 0.999 | 0.000 | 0.994 | 1.000 |
| 6 |  | 0.971 | 0.000 | 1.000 | 0.029 | 0.990 | 0.971 |
| 7 |  | 0.992 | 0.000 | 1.000 | 0.008 | 1.000 | 0.992 |
| Weighted |  | 0.938 | 0.014 | 0.986 | 0.062 | 0.937 | 0.938 |

In stage 3, seven classifiers are used. Training and testing datasets of stage 2 are now divided into 7 separate training and testing subsets. Each of the 7 training and testing data subset pairs have associated subclass labels, which range from 1 to 35, belonging to corresponding stage 2 class labels. Table 13 presents the number of instances available in both training and testing data subsets for stage 3 classifiers.

In stage 3, the batch size is 10 for the Random Forest classifier as the test data subset size is further reduced following the processing as a result of classification during stage 2. Since the attack class 6 (DoS) has only 1 specified subclass (18) in the dataset, it was not necessary to build a classifier for it. Tables 14 through 19 present the confusion matrices for all 6 other subclass classifiers.

For class label 1, the confusion matrix is presented in Table 14: the classifier struggles to distinguish between attack subclasses 31 and 32. Attack subclass 31 is dependent on the random pressure measurements sent to the device and attack subclass 32 is dependent on the negative pressure readings sent to the device. The classifier struggles to detect these two attacks whenever the random pressure sent to the device is also negative and detects 31 as 32 or 32 as 31. Additionally, subclass 29 also has relatively poor detection rates as it is misclassified as subclass 31 or 32 while a good number of patterns belonging to attack subclasses 31 or 32 are also classified as 29. The classifier for attack class 1 achieves an accuracy rate of only 82.30%. The FNR is 17.7%, which is very high compared to other classifiers for the same metric. Performance of classifier for attack class 2, presented in Table 15, is the second worst with 87.79% accuracy and 524 incorrectly classified instances. All other classifiers performed at a much higher level, specifically 99% or better as shown in Tables 16 through 20.

Table 13. Instance Counts for Stage 3

| Stage 2 Label | Stage 3 Label | Training Count | Testing Count |
|---|---|---|---|
| **4** | 1 | 1221 | 571 |
| | 2 | 1015 | 445 |
| | 3 | 1126 | 574 |
| | 4 | 1277 | 655 |
| | 5 | 931 | 485 |
| | 6 | 1326 | 700 |
| | 7 | 997 | 515 |
| | 8 | 1186 | 612 |
| | 9 | 936 | 460 |
| | 10 | 955 | 519 |
| | 11 | 1206 | 628 |
| | 12 | 1374 | 698 |
| **3** | 13 | 1077 | 517 |
| | 14 | 1158 | 518 |
| | 15 | 1148 | 510 |
| | 16 | 1115 | 543 |
| | 17 | 963 | 451 |
| **6** | 18 | 1449 | 727 |
| **5** | 19 | 1089 | 545 |
| **7** | 20 | 474 | 192 |
| **5** | 21 | 1134 | 588 |
| | 22 | 1009 | 533 |
| **7** | 23 | 1355 | 693 |
| | 24 | 752 | 408 |
| **2** | 25 | 995 | 477 |
| | 26 | 1237 | 571 |
| | 27 | 1389 | 690 |
| | 28 | 1233 | 625 |
| **1** | 29 | 1276 | 580 |
| | 30 | 1414 | 706 |
| | 31 | 1268 | 638 |
| | 32 | 1264 | 607 |
| **2** | 33 | 1071 | 533 |
| | 34 | 1327 | 683 |
| | 35 | 1491 | 713 |

Table 14. Stage 3 Class Label 1 Confusion Matrix

| Classified as → | 29 | 30 | 31 | 32 |
|---|---|---|---|---|
| **29** | 400 | 7 | 86 | 87 |
| **30** | 0 | 706 | 0 | 0 |
| **31** | 52 | 0 | 527 | 59 |
| **32** | 74 | 0 | 83 | 450 |

Table 15. Stage 3 Class Label 2 Confusion Matrix

| Classified as → | 29 | 30 | 31 | 32 | 33 | 34 | 35 |
|---|---|---|---|---|---|---|---|
| **25** | 368 | 41 | 3 | 42 | 2 | 4 | 17 |
| **26** | 37 | 479 | 0 | 45 | 0 | 4 | 6 |
| **27** | 2 | 0 | 671 | 1 | 2 | 0 | 14 |
| **28** | 35 | 74 | 3 | 423 | 51 | 13 | 26 |
| **33** | 0 | 0 | 7 | 6 | 505 | 0 | 15 |
| **34** | 0 | 0 | 7 | 2 | 0 | 647 | 27 |
| **35** | 1 | 0 | 7 | 3 | 10 | 17 | 675 |

Table 16. Stage 3 Class Label 3 Confusion Matrix

| Classified as → | 13 | 14 | 15 | 16 | 17 |
|---|---|---|---|---|---|
| **13** | 513 | 1 | 1 | 0 | 2 |
| **14** | 0 | 516 | 1 | 0 | 1 |
| **15** | 1 | 1 | 506 | 1 | 1 |
| **16** | 0 | 1 | 1 | 540 | 1 |
| **17** | 0 | 2 | 1 | 0 | 448 |

Table 17. Stage 3 Class Label 4 Confusion Matrix

| Classified as → | 1 | 2 | 3 | 4 | 5 | 6 | 7 | 8 | 9 | 10 | 11 | 12 |
|---|---|---|---|---|---|---|---|---|---|---|---|---|
| **1** | 569 | 1 | 0 | 0 | 0 | 0 | 0 | 0 | 1 | 0 | 0 | 0 |
| **2** | 1 | 431 | 0 | 0 | 0 | 6 | 0 | 0 | 1 | 6 | 0 | 0 |
| **3** | 0 | 0 | 573 | 1 | 0 | 0 | 0 | 0 | 0 | 0 | 0 | 0 |
| **4** | 0 | 0 | 0 | 655 | 0 | 0 | 0 | 0 | 0 | 0 | 0 | 0 |
| **5** | 0 | 0 | 0 | 0 | 481 | 2 | 0 | 1 | 0 | 0 | 0 | 1 |
| **6** | 1 | 2 | 0 | 0 | 0 | 692 | 0 | 0 | 0 | 5 | 0 | 0 |
| **7** | 1 | 0 | 0 | 1 | 0 | 0 | 512 | 0 | 1 | 0 | 0 | 0 |
| **8** | 0 | 0 | 0 | 0 | 2 | 0 | 0 | 609 | 0 | 1 | 0 | 0 |
| **9** | 0 | 0 | 0 | 1 | 0 | 0 | 0 | 0 | 458 | 1 | 0 | 0 |
| **10** | 0 | 1 | 0 | 0 | 0 | 0 | 1 | 0 | 2 | 515 | 0 | 0 |
| **11** | 0 | 1 | 1 | 2 | 0 | 0 | 0 | 0 | 0 | 0 | 624 | 0 |
| **12** | 0 | 0 | 0 | 0 | 0 | 1 | 0 | 1 | 0 | 0 | 0 | 696 |

Table 18. Stage 3 Class Label 5 Confusion Matrix

| Classified as → | 19 | 21 | 22 |
|---|---|---|---|
| **19** | 544 | 1 | 0 |
| **21** | 0 | 588 | 0 |
| **22** | 0 | 0 | 533 |

Table 19. Stage 3 Class Label 7 Confusion Matrix

| Classified as → | 20 | 23 | 24 |
|---|---|---|---|
| **20** | 187 | 0 | 5 |
| **23** | 0 | 693 | 0 |
| **22** | 2 | 0 | 406 |

Table 20. Stage 3 Performance

| Classifier | Accuracy | Correct Instances | Incorrect Instances | TPR | FPR | TNR | FNR | Precision | Recall |
|---|---|---|---|---|---|---|---|---|---|
| 1 | 82.30% | 2083 | 448 | 0.823 | 0.057 | 0.943 | 0.177 | 0.822 | 0.823 |
| 2 | 87.79% | 3768 | 524 | 0.878 | 0.020 | 0.980 | 0.202 | 0.876 | 0.878 |
| 3 | 99.37% | 2523 | 16 | 0.994 | 0.002 | 0.998 | 0.006 | 0.994 | 0.994 |
| 4 | 99.32% | 6815 | 47 | 0.993 | 0.001 | 0.999 | 0.007 | 0.993 | 0.993 |
| 5 | 99.94% | 1665 | 1 | 0.999 | 0.000 | 1.000 | 0.001 | 0.999 | 0.999 |
| 6 | 100.00% | 727 | 0 | 1.000 | 0.000 | 1.000 | 0.000 | 1.000 | 1.000 |
| 7 | 99.47% | 1286 | 7 | 0.995 | 0.002 | 0.998 | 0.005 | 0.995 | 0.995 |

**Comparison with Studies Reported in Literature**

Studies on the same dataset in the literature do not attempt to perform classification for the 35-subclass case as they consider the classification problem as a 7-category case. For a meaningful comparison with the other studies in the literature on the same dataset, it is then necessary to calculate the combined performance of stages 1 and 2. The accuracy of combined stages 1 and 2 is 0.9379×0.9806= 92.06%. Precision and recall rates are 0.937×0.982=0.920 and 0.938×0.982=0.921, respectively.

Several studies using the same gas pipeline dataset were reported in the literature as presented in Table 21, which considered the classification problem for seven classes. In Table 21, et Perez et al. [18] reports better performance than the current study. Their approach employed 80-20% ratio for splitting the dataset into training and testing subsets, which is not the same as the 67-33% split ratio for the dataset as is typical for most studies including the current one. Demertzis et al. [14] proposed one-class anomaly detection approach for this dataset. Apart from being an anomaly (versus misuse) detection system, their study exposes several important differences when compared to ours. One significant difference is that they did not employ the full dataset in their study. They subsampled 97,019 instances from 274,628 instances in the original dataset. In light of this, one can argue that the two studies are not directly comparable. For the study reported in [19], results are poor and well below what the current study achieved for the set of comparable performance metric values.

Table 21. Performance Comparison with Other Studies in Literature

| Classifier | Accuracy | Precision | Recall |
|---|---|---|---|
| This Study | 92.06% | 0.920 | 0.921 |
| Random Forest [18] | 99.41% | 0.994 | 0.994 |
| K-Means [19] | 56.80% | 0.832 | 0.573 |
| GMM [19] | 45.16% | 0.731 | 0.442 |
| OCC-eSNN [14] | 98.82% | 0.988 | 0.988 |
| OCC-SVM [14] | 97.98% | 0.980 | 0.980 |

## Conclusions

In this study, we proposed a three-stage classifier for detecting and identifying known intrusions for a gas pipeline-based SCADA network. The dataset, which was developed using an academic laboratory setup at the Mississippi State University, entails 3 attack groups, 7 attack classes and 35 attack subclasses. Random Forest classifier was used for all stages. Simulation results showed that 24 out of 35 attack subclasses, which belonged to attack classes 3, 4, 5, 6, and 7, were detected and identified with relatively high accuracy while the performance for the remaining 11 attack subclasses, which were associated with attack classes 1 and 2, was not at par as it lagged by a considerable margin. In light of the fact that studies in literature which attempt to detect and identify at the level of tens of attack subclasses are scarce at best, the proposed and novel three-stage classifier model in this study is promising for its overall performance to address such problems.